\documentclass[11pt,prd,preprint,superscriptaddress]{revtex4}
\usepackage{revsymb}
\usepackage{graphicx}
\usepackage{bm}
\usepackage{amsmath}
\usepackage{amssymb}
\usepackage{graphicx}
\begin{document}

\title{TESTING THE RASTALL'S THEORY USING\\
MATTER POWER SPECTRUM}

\author{C.E.M.Batista\footnote{E-mail: cedumagalhaes22@hotmail.com}, J.C. Fabris\footnote{E-mail: fabris@cce.ufes.br} and M. Hamani Daouda\footnote{E-mail: daoudah8@yahoo.fr}}
\affiliation{Universidade Federal do Esp\'{\i}rito Santo,
Departamento
de F\'{\i}sica\\
Av. Fernando Ferrari, 514, Campus de Goiabeiras, CEP 29075-910,
Vit\'oria, Esp\'{\i}rito Santo, Brazil}

\begin{abstract}
The Rastall's theory is a modification of the General Relativity theory leading to
a different expression for the conservation law in the matter sector compared with
the usual one. It has been argued recently that such a theory may have applications
to the dark energy problem, since a pressureless fluid may lead to an accelerated
universe. In the present work we confront the Rastall's theory with the
power spectrum data. The results indicate a configuration that essentially reduces
the Rastall's theory to General Relativity, unless the non-usual conservation law
refers to a scalar field, situation where other configurations are eventually possible.
\end{abstract}

\maketitle
\date{\today}

\section{Introduction}

A large number of cosmological observational data requires the existence of
two exotic components in the matter content of the universe, dark matter and dark energy
\cite{kamion}.
Both constitute the so-called dark sector of the cosmic budget. Dark matter is
necessary, for example, to explain the formation of structures in the expanding universe.
Dark energy is necessary, on the other hand to explain the present stage of the
accelerated expansion of the universe. Dark energy must have a negative pressure in order
to induce the acceleration of the universe. This is a quite strange property, and
one of the most important question today in theoretical physics concerns the nature
of such an exotic fluid. All candidates to describe the dark energy component face serious problems
and drawback.
\par
Another possibility to explain the observational data is to consider that General
Relativity is not the true gravitational theory. Some modifications of General
Relativity may lead to an accelerated universe even if only usual types of matter
are taked into account. A very fashion proposal is the $f(R)$ theories, a non-linear
generalization of the Einstein-Hilbert action . For a review of the $f(R)$ theories and their present status, see
reference \cite{felice}. In fact, the unusual properties of dark energy motivate the search of
many other alternatives to explain the observational data.
\par
Recently, it has been investigated if the Rastall's theory may be a viable alternative to the introduction of
dark energy in the General Relativity context \cite{capone}. The Rastall's theory is
a modification of the General Relativity, proposed in 1972 \cite{rastall}. It implies a change
of the Einstein's equation that mounts out to a modification of the conservation law for
the energy-momentum tensor. One of the motivations to this proposal is the fact that
the usual conservation law is only firmly tested in special relativity. The theory contains a free parameter $\lambda$ such that
$\lambda = 1$ implies General Relativity. In this way, it can be considered as a deformation
of General Relativity. The modification in the conservation law can lead to new effects
compared with General Relativity. Depending on the value of the parameter $\lambda$,
a pressureless matter, for example, can induce an accelerated expansion.
\par
An important drawback of
Rastall's theory is the absence of the Langrangian formulation. But, it is possible
that it can obtained from an action principle using Weyl's geometry \cite{smalley}.
It has been argued in reference \cite{lee} that the Rastall's theory may appear even in the context of
Riemannian geometries by a redefinition of the energy-momentum tensor. In any case,
the Rastall's theory implies that a given fluid, with a specific equation of state, may have a very different effective
equation of state when interpreted in the context of General Relativity. This properties
has called the attention to this proposal as an alternative to dark energy.
\par
In reference \cite{capone}, the Rastall's theory has been confronted agains the
supernova type Ia data. The result indicates that it is competitive with respect
to the $\Lambda$CDM model, but the mass density is very high, around $\Omega_0 \sim 0.5$. This suggests
that the theory may face troubles when tested against the mass agglomeration phenomena.
\par
In the present work, we intend to test the Rastall's theory using the power spectrum observational data. This implies a perturbative study. A perturbative study of
the Rastall's theory has already been performed in reference \cite{kerner}. The main
conclusion was that the theory behaves also at perturbative level like General Relativity
but with an effective equation of state. In particular, if the fluid has an effective
equation of state necessary to induce the accelerated expansion of the universe, it will
present a negative squared effective sound speed. Hence, the resulting scenario is plagued with
instabilities. Moreover, in order to compare the theoretical predictions of the Rastall's
theory with the power spectrum data, it seems unavoidable to introduce a two fluid model.
This represents a problem in the original Rastall's theory. A possible way out is to consider
a fluid which obeys the conservation law of Rastall's theory, with another one that
conserves separately in the usual way. This may be justified if the former fluid is in fact
a field, like a scalar field, that obeys a modified Klein-Gordon equation.
In fact, if both fluids have a hydrodynamical representation, the power spectrum predicts
$\lambda = 1$ as it will be described later in this work - that is, the theory reduces to General Relativity. If the Rastall's fluid
is a scalar field, the situation is more complex, and even if $\lambda = 1$ remains favored,
other possibility appears.
\par
This paper is organized as follows. In next section, we present the Rastall's theory, deriving some cosmological relations. In section III, the matter power spectrum is
determined in the hydrodynamical representation. In section IV, the matter power spectrum
is determined for the case where one of the fluids is representing by a self-interacting
scalar field. In section V we present our conclusions.
\par

\section{The field equations}

Originally, the fundamental equations of the Rastall's theory were written as
\begin{eqnarray}
\label{re1}
 R_{\mu\nu} - \frac{\lambda}{2}g_{\mu\nu}R &=& \kappa T_{\mu\nu},\\
\label{re2}
{T^{\mu\nu}}_{;\mu} &=& \frac{1 - \lambda}{2\kappa}R^{;\nu}.
\end{eqnarray}
The General Relativity theory, with the usual conservation of the energy-momentum tensor,
is re-obtained when $\lambda = 1$. These equations may be re-written as
\begin{eqnarray}
\label{re1bis}
 R_{\mu\nu} - \frac{1}{2}g_{\mu\nu}R &=& \kappa\biggr\{T_{\mu\nu} - \frac{\gamma - 1}{2}g_{\mu\nu}T\biggl\},\\
\label{re2bis}
{T^{\mu\nu}}_{;\mu} &=& \frac{\gamma - 1}{2}T^{;\nu},
\end{eqnarray}
where
\begin{equation}
 \gamma = \frac{3\lambda - 2}{2\lambda - 1}.
\end{equation}
Again, $\gamma = 1$ (corresponding to $\lambda = 1$) implies General Relativity.
Equivalently, one can recast these equations under the following form:
\begin{eqnarray}
\label{e1}
 R_{\mu\nu} &=& \kappa\biggr\{T_{\mu\nu} - \frac{2 - \gamma}{2}g_{\mu\nu}T\biggl\},\\
\label{e2}
{T^{\mu\nu}}_{;\mu} &=& \frac{\gamma - 1}{2}T^{;\nu},
\end{eqnarray}
\par
In analysing the perturbed Rastall model, we must consider a two fluid model. The reason
is that baryons clearly exists, and there are good observational evidences that baryons
can be modelled by a pressureless matter with an approximately zero sound velocity. This indicates that baryons can not be introduced directly in the framework of equations
(\ref{re1},\ref{re2}) by simply adding a new energy-momentum tensor in the rhs of (\ref{re1}): if the original framework of the Rastall's theory is preserved, even if we set a fluid with zero pressure the resulting sound velocity is not zero.
Alternatively, a conserved baryonic energy-momentum tensor could be added to the rhs
of (\ref{re1}), but this would imply a presence of an inhomogeneous term for the exotic
fluid obeying (\ref{re2}). This inhomogeneous term leads to a negative energy component
of the exotic fluid, which dominates either in the past or in the future. For this
reason, we will consider equations (\ref{re1bis},\ref{re2bis}) the fundamental framework of the two fluid model including baryons; it implies the addition of a second energy-momentum tensor
which obeys the usual conservation law. This additional energy-momentum tensor will
represent the baryonic component.
\par
Hence, we will consider a cosmological model with two fluids, one obeying
the Rastall's framework, with no usual conservation of the corresponding energy-momentum tensor,
and the other obeying the traditional conservation law of general relativity.
Under these conditions, the field equations read,
\begin{eqnarray}
\label{fe1}
 R_{\mu\nu} &=& \kappa\biggr\{T^x_{\mu\nu} - \frac{2 - \gamma}{2}g_{\mu\nu}T^x\biggl\}
+ \kappa T^m_{\mu\nu},\\
\label{fe2}
{T_x^{\mu\nu}}_{;\mu} &=& \frac{\gamma - 1}{2}T_x^{;\nu},\\
\label{fe3}
{T_m^{\mu\nu}}_{;\mu} &=& 0.
\end{eqnarray}
The subscripts (superscripts) $x$ and $m$ designate the exotic and matter (baryonic) components,
respectively.
\par
The universe is homogenous and isotropic at least at scales greater than
$100\,Mpc$. Hence, at sufficiently large scales, it can be represented by
the Friedmann-Lema\^{\i}tre-Robertson-Walker (FLRW) metric,
\begin{equation}
 ds^2 = dt^2 -
a(t)^2\biggr\{\frac{dr^2}{1 - kr^2} + r^2(d\theta^2 +\sin^2\theta d\phi^2)\biggl\},
\end{equation}
where $a(t)$ is the scale factor and $k$ is the curvature of the spatial section.
Introducing this metric in the field equations (\ref{fe1},\ref{fe2},\ref{fe3}), and
specializing for the flat case ($k = 0$), we obtain the following equations of motion:
\begin{eqnarray}
\biggr(\frac{\dot a}{a}\biggl)^2 &=& \frac{\kappa}{6}\biggr\{3 - \gamma - 3 (1 - \gamma)\omega_x\biggl\}\rho_x + \frac{\kappa}{3}\rho_m,\\
\frac{\ddot a}{a} &=& - \frac{\kappa}{6}\biggr\{\gamma + 3(2 - \gamma)\omega_x\biggl\}\rho_x
- \frac{\kappa}{6}\rho_m,\\
\dot\rho_x + 3\frac{\dot a}{a}(1 + \omega_x)\rho_x &=& \frac{\gamma - 1}{2}(1 - 3\omega_x)\dot\rho_x,\\
\dot\rho_m + 3\frac{\dot a}{a}\rho_m &=& 0.
\end{eqnarray}
The two last equation can be integrated leading to,
\begin{eqnarray}
 \rho_x &=& \rho_{x0}a^\frac{- 6(1 + \omega_x)}{3 - \gamma + 3(\gamma - 1)\omega_x},\\
\rho_m &=& \rho_{m0}a^{-3}.
\end{eqnarray}
Remark that all this formulation is equivalent to a traditional General Relativity,
where the perfect fluid would have an effective equation of state given by,
\begin{equation}
\label{efetivo}
 \omega_{eff} = \frac{\gamma - 1 + (5 - 3\gamma)\omega_x}{3 - \gamma + 3(\gamma - 1)\omega_x}.
\end{equation}
It is interesting that $\omega_x = - 1$ implies $\omega_{eff} = - 1$: the cosmological constant case is a fix point in this stuff.
The reason for this fixed point is easily understood inspecting (\ref{re2}): it corresponds to a de Sitter (or anti-de Sitter) space-time,
whose curvature is constant; consequently, the usual conservation law is recovered.
\par
Using the modified the flat Friedmann's equation, and defining, as usual,
\begin{equation}
 \Omega_{x0} = \frac{\kappa\rho_{x0}}{3H^2_0} \quad , \quad \Omega_{0} = \frac{\kappa\rho_{m0}}{3H^2_0}\quad ,
\end{equation}
the following equation must be obeyed today:
\begin{equation}
\label{rel1}
1 =  \frac{\Omega_{x0}}{2}\biggr\{3 - \gamma - 3 (1 - \gamma)\omega_x\biggl\} + \Omega_{0}.
\end{equation}
Remark that to define the density parameters the newtonian cosmological constant is employed. It could be used an effective cosmological constant,
leading to a numerical different value for the density parameter, but without changing the general framework \cite{capone}.

\section{Perturbative analysis: the fluid description}

Let us consider in Rastall's theory the behaviour of a given fluid
characterized, for example, by an equation of state $p = \omega\rho$, with $\omega$ constant.
The predictions of the Rastall's theory, in one fluid description, is equivalent for
the background point of view, to the predictions of General Relativity when a fluid
with an equation of state of the type $p = \omega_{eff}\rho$, $\omega$ and $\omega_{eff}$ being
connected by the relation (\ref{efetivo}). In this sense, with this identification, one
theory can be mapped into the other.
\par
At perturbative level, however, this equivalence is not evident. In reference (\cite{kerner}), a perturbative study has been carried out, considering just one fluid,
obeying the framework of the Rastall's theory. The fluid description was kept all along
the calculation. The final equations reveal the same results of General Relativity, provided
that the identification (\ref{efetivo}) is made. The equivalence remains at perturbative
level as far as the fluid description is used.
\par
One of the most important problems today
in cosmology is the description of dark energy. Dark energy requires negative pressure
and, at perturbative level, a fluid with negative pressure is unstable at small scales
\cite{jerome}. The Rastall's theory opens new possibility, but as far as we remain at
the level of a fluid description, and with a one-fluid model, it seems that the same problems remain: a negative effective pressure is required at background level, and this leads
to instabilities at perturbative level using the results of reference \cite{kerner}.
\par
But, as already briefly discussed in the Introduction and in the previous section, we can go one step further by considering the two fluid model, one of them
representing the baryons. This allows to consider the observational data to restrict the
model. In fact, the power spectrum concerns the baryonic component. In other words,
it concerns a fluid with zero effective pressure, what assures the gravitational collapse,
leading to the formation of local structures. The other fluid will follow the
Rastall's framework. Its considered as the dark component of the universe. 
To be specific, this exotic fluid will be taken having zero pressure, leading to
an effective equation of state parameter
\begin{equation}
 \omega_{eff} = \frac{\gamma - 1}{3 - \gamma}.
\end{equation}
In this situation, acceleration of the universe can be achieved if
$\gamma < 0 $ or $\gamma > 3$. Moreover, relation (\ref{rel1}) reads now,
\begin{equation}
\label{rel1bis}
1 =  \Omega_{x0}\frac{3 - \gamma}{2} + \Omega_{0}.
\end{equation}
\par
Let us now consider the perturbation of the two fluid models. We will work in
the synchronous gauge. Since the modes of interest for the matter power spectrum
are well inside the horizon, the results are not sensitive to the use of one gauge or
another, or even a gauge invariant formalism.
\par
In the synchronous gauge condition, we introduce the perturbations in the metric
and matter functions,
\begin{eqnarray}
\label{ep1}
 \rho_x = \rho_{x0} + \delta\rho_x \quad , \quad \rho_m = \rho_{m0} &+& \delta\rho_m\quad ,
\quad u_m = u_{m0} + \delta u_m\quad ,\\
\label{ep2}
p_x = p_{x0} + \delta p_x \quad , \quad p_m = p_{m0} &+& \delta p_m\quad ,\quad u_x = u_{x0} + \delta u_x\quad , \\
\label{ep3}
g_{\mu\nu} = g^0_{\mu\nu} + h_{\mu\nu} \quad &,& \quad h_{\mu0} = 0.
\end{eqnarray}
In expressions (\ref{ep1}-\ref{ep3}), the sub(super)scripts ``0'' indicate the background
functions, and $\delta\rho_m$, $\delta\rho_x$, $\delta p_m$, $\delta p_x$, $\delta u_m$, $\delta u_x$, $h_{\mu\nu}$ indicate the
perturbed quantities in density, pressure, four-velocity and metric, while $h_{\mu0} = 0$ defines the coordinate condition.
Since both fluids have zero pressure, we fix $p_x = p_m = 0$. Moreover, since
we will not taken into account entropic perturbations (which nevertheless may lead
to new interesting effects in the present framework) we can fix $\delta p_x = \delta p_m = 0$.
\par
Introducing these perturbations and remaining at the linear level, we obtain the following
coupled equations for the perturbed quantities:
\begin{eqnarray}
\label{p1}
 \ddot\delta_m + 2\frac{\dot a}{a}\dot\delta_m &=& \frac{3}{2}\gamma\Omega_{x0}a^{- \frac{6}{3 -\gamma}}\delta_x + \frac{3}{2}\Omega_{0}a^{-3}\delta_m,\\
\label{p2}
\dot\delta_x &=& - \frac{2}{3 - \gamma}(\theta_x - \dot\delta_m),\\
\label{p3}
\dot\theta_x + \frac{9 - 5\gamma}{3 - \gamma}\frac{\dot a}{a}\theta_x &=& - \frac{1 - \gamma}{2}k^2\frac{\delta_x}{a^2},\\
\label{p4}
\delta_m &=& \frac{h}{2}.
\end{eqnarray}
In these expressions, we have the following definitions:
\begin{eqnarray}
 \delta_x = \frac{\delta\rho_x}{\rho_x} \quad , \quad \delta_m = \frac{\delta\rho_m}{\rho_m}
\quad , \quad \theta = \partial_k\delta u_k \quad , \quad h = \frac{\sum_{k=1}^3h_{kk}}{a^2}
\quad .
\end{eqnarray}
\par
In order to integrate numerically perturbed equations (\ref{p1}-\ref{p3}) it is more convenient to
use the scale factor $a$ and as the dynamical variable, since it is directly connected
with the dimensionless redshift quantity through $z = - 1 + \frac{1}{a}$ (fixing today $a_0 = 1$).
In terms of this new variable, we obtain the following equations:
\begin{eqnarray}
\label{p1bis}
 {\delta''}_m + \biggr(\frac{2}{a} + \frac{f'(a)}{f(a)}\biggl){\delta'}_m &=& \frac{3}{2}\gamma\frac{\Omega_{x0}}{f^2(a)}a^{- \frac{6}{3 -\gamma}}\delta_x + \frac{3}{2}\frac{\Omega_{0}}{f^2(a)}a^{-3}\delta_m,\\
\label{p2bis}
\delta'_x &=& - \frac{2}{3 - \gamma}\biggr(\frac{\theta_x}{f(a)} - \delta'_m\biggl),\\
\label{p3bis}
\theta'_x + \frac{9 - 5\gamma}{3 - \gamma}\frac{\theta_x}{a} &=& - \frac{1 - \gamma}{2}\frac{k^2}{k_0^2}\frac{\delta_x}{a^2f(a)},\\
\end{eqnarray}
where $k_0$ is the wavenumber associated to the Hubble's radius, and it has been defined
the function,
\begin{equation}
 f(a) = \biggr[\frac{3 - \gamma}{2}\Omega_{x0}a^{\frac{2\gamma}{\gamma - 3}} + \frac{\Omega_{0}}{a}\biggl]^{1/2}.
\end{equation}
\par
To obtain a prediction, we compare the matter power spectrum, defined by,
\begin{equation}
 {\cal P} = \delta_k^2,
\end{equation}
with the observational data of the 2dFGRS observational program \cite{cole}. To fix the initial condition
we use the BBKS transfer function \cite{sugi,bardeen}, and apply the prescription described in the
reference \cite{sola}.
We use the statistical $\chi^2$ parameter defined
by
\begin{equation}
 \chi^2 = \sum_{i=1}^n\frac{({{\cal P}^i}_k^o -  {{\cal P}^i}_k^t)^2}{\sigma_i^2},
\end{equation}
where ${{\cal P}^i}_k^o$ is the $ith$ observational data, $\sigma_i^2$ being the
error bar, ${{\cal P}^i}_k^t$ is the
corresponding theoretical prediction. The probability distribution function (PDF) is given
by
\begin{equation}
 P(\gamma,\Omega_{0}) = Ae^{-\chi^2/2},
\end{equation}
where $A$ is a normalization constant. As indicated, the PDF depends on two free
parameters, the matter density $\Omega_0$ and the $\gamma$ parameter which characterizes
the deviation from the Einstein theory.
Marginalizing (integrating) in one of the parameters, we obtain the one-dimensional PDF
for the remaining free parameter.
The results are shown in figure 1. The probability is highly concentrated around
$\gamma = 1$, which corresponds to the Einstein theory - it admits a slight deviation
if the matter parameter is high, almost without the exotic fluid. This result can
be easily understand: as far as the parameter $\gamma$ differs from the General Relativity
value $\gamma = 1$, oscillations or even exponential behaviour (depending if
$\gamma$ is lesser or greater than 1, respectively) is induced in the exotic fluid,
and this is highly transfered to the matter fluid. Such behaviour is not observed in
the matter spectrum. Remark that if $\Omega_{x0} \sim 0$, implying $\Omega_0 \sim 1$, larger
deviations from $\gamma = 1$ are possible.
\par
Hence, we can conclude that the value $\gamma = 1$ is the prediction obtained from
the matter power spectrum, with a precision of the order of $10^{-5}$ (see the
graphics). In fact, the best fitting is achieved by $\gamma = 1$ and $\Omega_0 \sim 0.79$,
$\chi^2 = 0.38$ per degree of freedom. The best fit model $\Lambda$CDM model has the same
$\chi^2$ per degree of freedom. 

\begin{center}
\begin{figure}[!t]
\begin{minipage}[t]{0.31\linewidth}
\includegraphics[width=\linewidth]{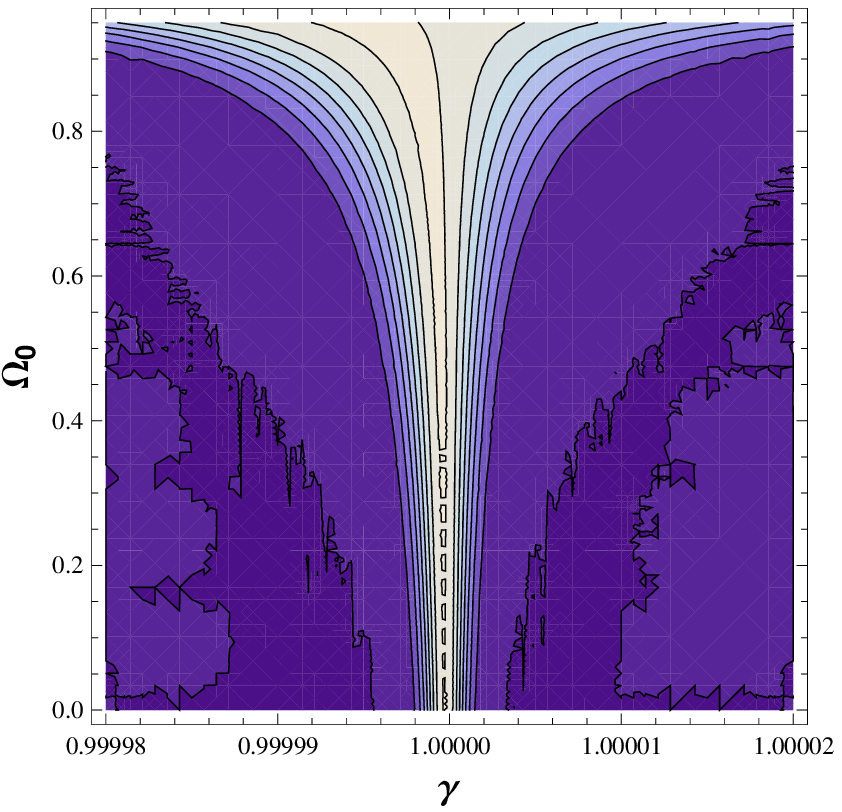}
\end{minipage} \hfill
\begin{minipage}[t]{0.33\linewidth}
\includegraphics[width=\linewidth]{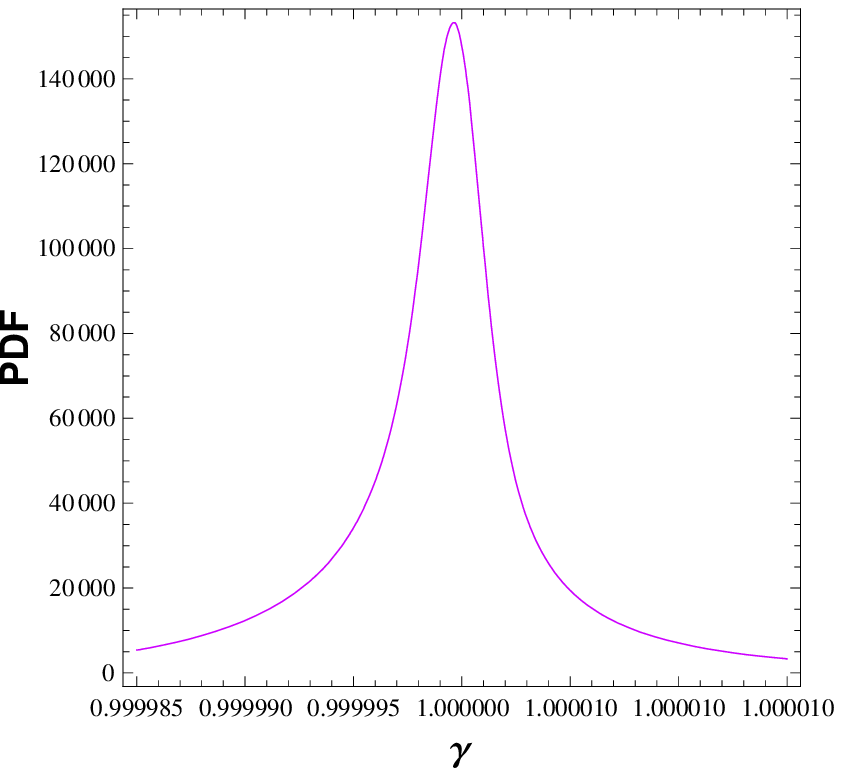}
\end{minipage} \hfill
\begin{minipage}[t]{0.3\linewidth}
\includegraphics[width=\linewidth]{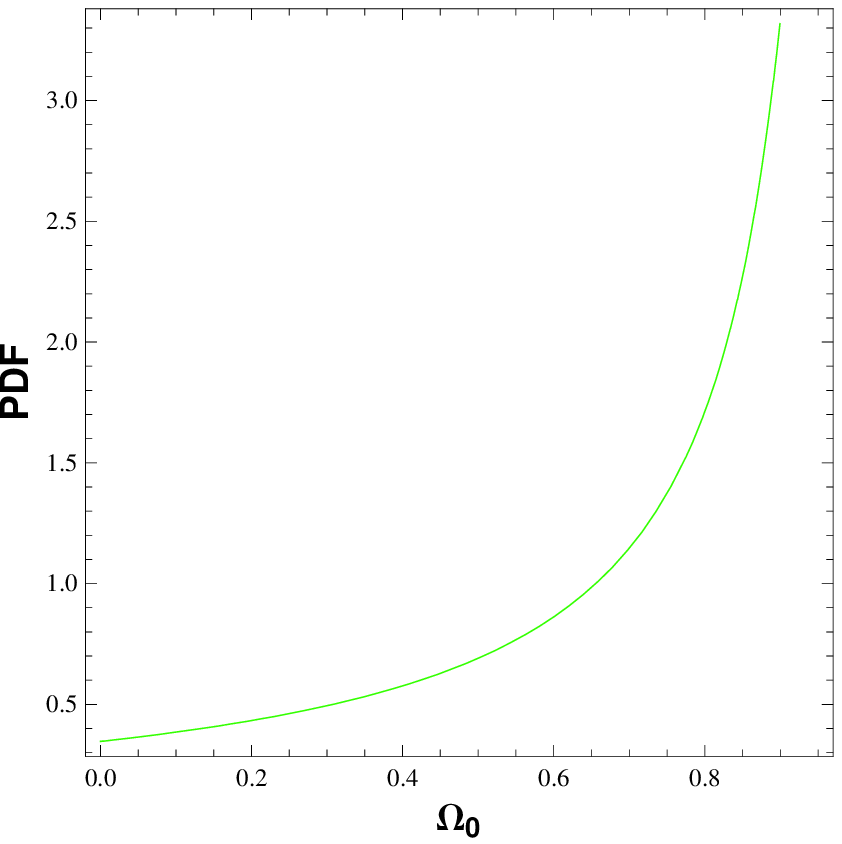}
\end{minipage} \hfill
\caption{{\protect\footnotesize On the left, the two-dimensional probability distribution
function (PDF) for the matter density $\Omega_0$ and the $\gamma$ parameter. In the center the
one-dimensional PDF for the $\gamma$ parameter, and on the right the one-dimensional
PDF for the matter density parameter $\Omega_0$.}}
\end{figure}
\end{center}

\section{Perturbative analysis: the scalar field description}

The result described above is restricted to a fluid formulation of both components,
that representing matter and that representing the exotic fluid. This exotic fluid
must, more precisely, be represented by a field than a fluid - it is very difficult that
a usual fluid could present the exotic behaviours connected with dark matter and dark energy, or even follow the new conservation
law dictated by the Rastall's theory.
In this sense, let us consider the most simple field description in cosmology, that of a self-interacting
scalar field. The energy momentum tensor of this field is given by,
\begin{equation}
 T_{\mu\nu} = \phi_{;\mu}\phi_{;\nu} - \frac{1}{2}g_{\mu\nu}\phi^{;\rho}\phi_{;\rho} +
g_{\mu\nu}V(\phi).
\end{equation}
If we impose the usual conservation law for the energy-momentum tensor ${T^{\mu\nu}}_{;\mu} = 0$, we obtain the usual Klein-Gordon equation:
\begin{equation}
 \nabla_\rho\nabla^\rho\phi = - V_{\phi}(\phi).
\end{equation}
However, if now the conservation law must read as in equation (\ref{e2}), the scalar
field must now obey the following equation:
\begin{equation}
 \nabla_\rho\nabla^\rho\phi + (3 - 2\gamma)V_\phi = (1 - \gamma)\frac{\phi^{;\rho}\phi^{;\sigma}\phi_{;\rho;\sigma}}{\phi_{;\alpha}\phi^{;\alpha}}.
\end{equation}
In the two fluid model, the corresponding ``Einstein's'' equation takes the form,
\begin{equation}
 R_{\mu\nu} - \frac{1}{2}g_{\mu\nu}R = \phi_{;\mu}\phi_{;\nu} - \frac{2 - \gamma}{2}g_{\mu\nu}\phi_{;\rho}\phi^{;\rho} + g_{\mu\nu}(3 - 2\gamma )V(\phi).
\end{equation}
From the scalar field description it comes out some similarities of the Rastall's
theory with the K-Essency models \cite{mukhanov}. K-Essence models may also be plagued with negative
sound velocity problems what strength the mentioned similarities \cite{linder}.
\par
One clear advantage in using the scalar field representation is that it usually avoids
the problems of the fluid representation of components with negative pressure.
However, while this is clear in the usual case, with the Einstein's equation coupled to the
ordinary Klein-Gordon's equation, this is less clear in the framework of the Rastall's
theory. In order to investigate the constraints from mass power spectrum for this
field formulation of the Rastall's theory, we take the coupled system matter+scalar field+Rastall's gravity, written in a convenient way for the perturbative analysis in
the synchronous coordinate condition:
\begin{eqnarray}
\label{fe1bis}
 R_{\mu\nu} &=& \phi_{;\mu}\phi_{;\nu} + \frac{1 - \gamma}{2}g_{\mu\nu}\phi_{;\rho}\phi^{;\rho} - (3 - 2\gamma)g_{\mu\nu}V + 8\pi G\biggr(T_{\mu\nu} - \frac{1}{2}g_{\mu\nu}T\biggl),\\
\label{fe2bis}
\nabla_\rho\nabla^\rho\phi &=& - (3 - 2\gamma)V_\phi + (1 - \gamma)\frac{\phi^{;\rho}\phi^{;\sigma}\phi_{;\rho;\sigma}}{\phi_{;\alpha}\phi^{;\alpha}},\\
\label{fe3bis}
{T^{\mu\nu}}_{;\mu} &=& 0.
\end{eqnarray}
\par
Now, let us suppose that the scalar field has zero pressure.
Since the energy and the pressure of the scalar field is given by
\begin{eqnarray}
 \rho_\phi &=& \frac{\dot\phi^2}{2} + V(\phi),\\
p_\phi &=& \frac{\dot\phi^2}{2} - V(\phi), 
\end{eqnarray}
the condition of zero pressure implies $\dot\phi^2/2 = V$, leading to $\rho_\phi = \dot\phi^2$.
From the point of view of the background, there is no difference between the fluid and scalar field approach. But at perturbative level the difference is significative. 
\par
In the synchronous coordinate condition, the perturbed equations corresponding to
the system described by (\ref{fe1bis}-\ref{fe3bis}), are the following:
\begin{eqnarray}
 \ddot\delta + 2\frac{\dot a}{a}\dot\delta - \frac{3}{2}\frac{\Omega_0}{a^3}\delta &=&
(3 - \gamma)\dot\phi\dot\Psi - (3 - 2 \gamma)V_\phi\Psi,\\
\gamma\ddot\Psi + 3\frac{\dot a}{a}\dot\Psi + \biggr\{\frac{k^2}{a^2} + (3 - 2\gamma)V_{\phi\phi}\biggl\}\Psi &=& \dot\phi\dot\delta,
\end{eqnarray}
where $\Psi = \delta\phi$ and $\delta$ is the density contrast of the matter component.
Using now the scale factor as the variable, the above system of equations take the
following form:
\begin{eqnarray}
 \delta'' + \biggr\{\frac{2}{a} + \frac{f'(a)}{f(a)}\biggl\} \dot\delta - \frac{3}{2}\frac{\Omega_0}{a^3f^2(a)}\delta &=&
(3 - \gamma)\phi'\Psi' - (3 - 2 \gamma)\frac{V_\phi}{f^2(a)}\Psi,\\
\gamma\Psi'' + \biggr\{\frac{3}{a} + \gamma\frac{f'(a)}{f(a)}\biggl\}\Psi' + \biggr\{\frac{k^2}{a^2f^2(a)} + (3 - 2\gamma)\frac{V_{\phi\phi}}{f^2(a)}\biggl\}\Psi &=& \phi'\delta',
\end{eqnarray}
where
\begin{eqnarray}
 \phi' &=& \sqrt{3\Omega_{x0}}\frac{a^\frac{-3}{3 - \gamma}}{f(a)},\\
V &=& \frac{3}{2}\Omega_{x0}a^\frac{-6}{3 - \gamma},\\
V_\phi &=& \frac{V_a}{\phi'},\\
V_{\phi\phi} &=& \frac{1}{\phi'}\frac{d}{da}V_\phi.
\end{eqnarray}
The function $f(a)$ is defined as before.
\par
We perform the same statistical analysis as before, again imposing the initial
conditions using the BBKS transfer function. 
The results are shown in figure 2. The main difference is that there is now two relevant
regions in the space parameter: one around $\gamma = 1$, with a low density, and the
other near $\gamma = 0$, but positive, extending from the low to high density. The
region around $\gamma = 1$ has high probabilities, but it is smaller; the region
near $\gamma = 0$ has lower probabilities, but extend to a large region.
The consequence is that, after marginalization, there is two peaks in the one-dimensional
probability for $\gamma$: one near $\gamma = 1$ and another near $\gamma = 0$. The second
peak is higher. We think that this is an effect of the larger probability region around
$\gamma = 0$, which seems to compensate the higher probabilities around $\gamma = 1$.
There are also two peaks in the one-dimensional PDF for $\Omega_0$, one near $\Omega_0 = 0$ and another near $\Omega_0 = 1$. The best fitting is achieved for $\gamma = 1.02$ and
$\Omega_0 = 0.72$ with a $\chi^2$ per degree of freedom equal $0.30$ better than the
$\Lambda$CDM model. The PDF goes quickly to zero for $\gamma < 0$. Still concerning the
peak near $\gamma = 0$, the parameters for $\gamma$ and $\Omega_0$ in this region
implies a $\chi^2$ around 0.33, high compared with the $\chi^2$ for $\gamma \sim 1$, but
smaller than the $\Lambda$CDM best fitting. We may ask about
the statistical relevance of this second peak.

\begin{center}
\begin{figure}[!t]
\begin{minipage}[t]{0.3\linewidth}
\includegraphics[width=\linewidth]{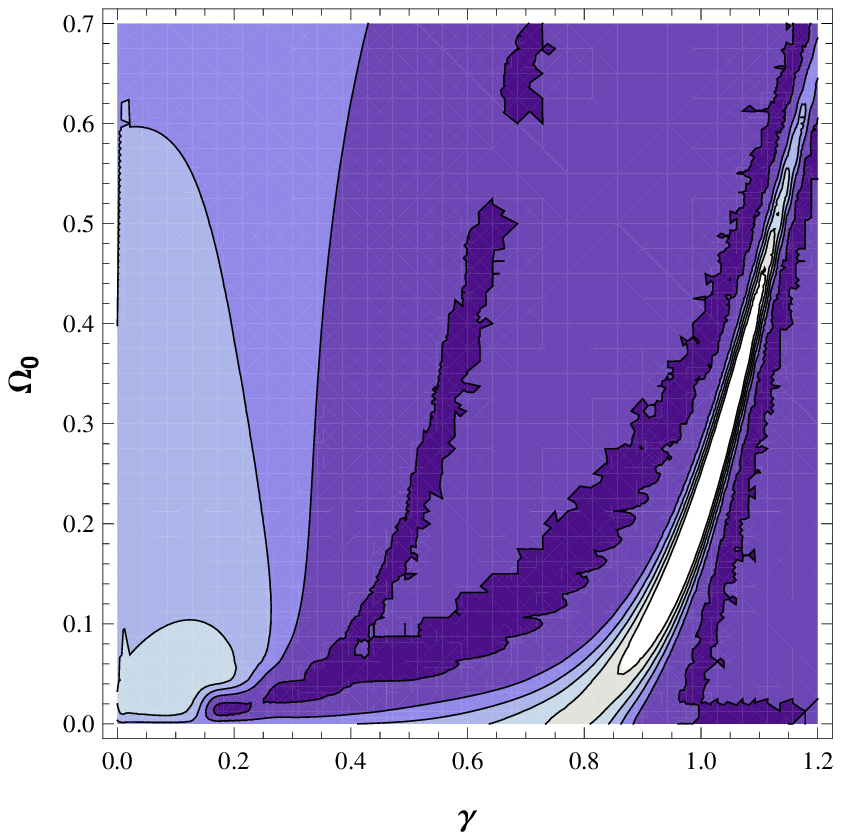}
\end{minipage} \hfill
\begin{minipage}[t]{0.3\linewidth}
\includegraphics[width=\linewidth]{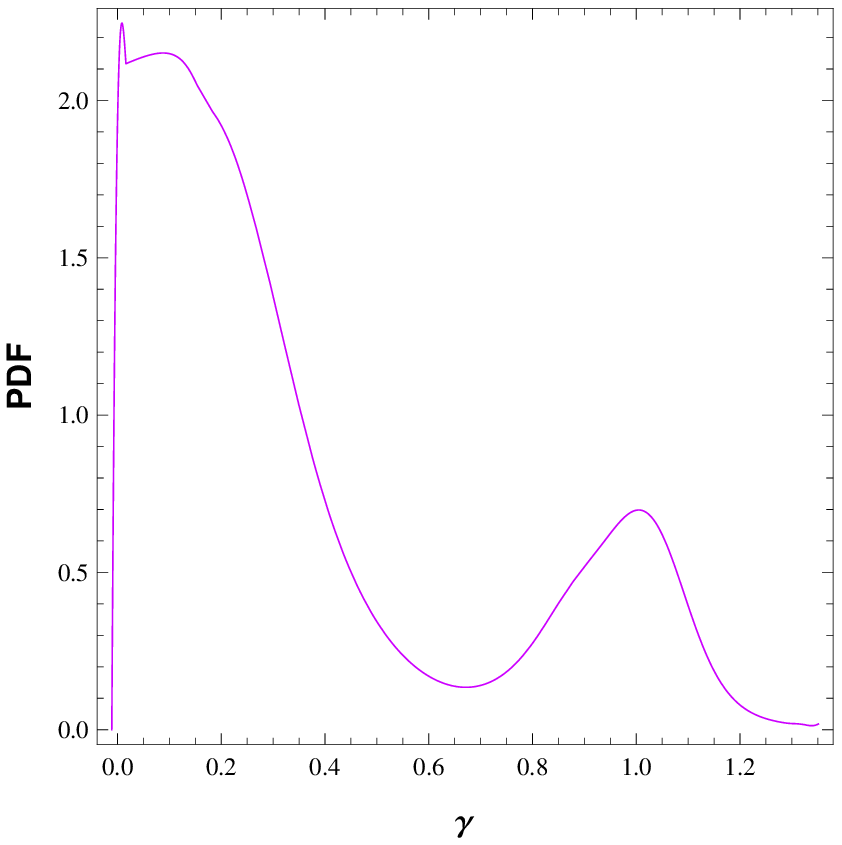}
\end{minipage} \hfill
\begin{minipage}[t]{0.3\linewidth}
\includegraphics[width=\linewidth]{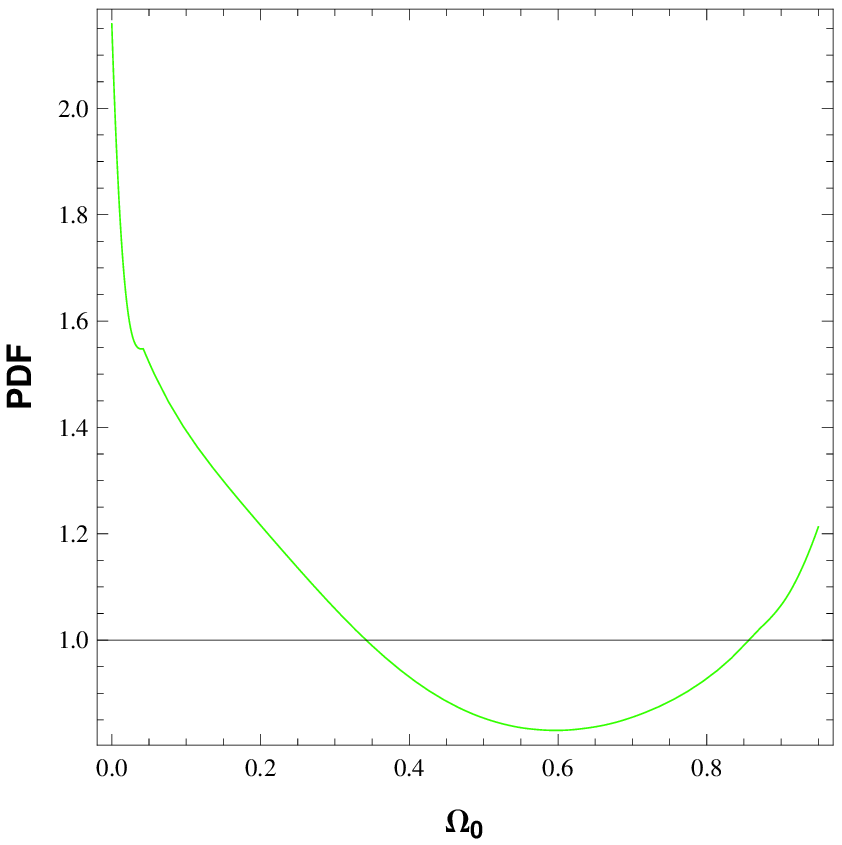}
\end{minipage} \hfill
\caption{{\protect\footnotesize On the left, the two-dimensional probability distribution
function (PDF) for the matter density $\Omega_0$ and the $\gamma$ parameter. In the center the
one-dimensional PDF for the $\gamma$ parameter, and on the right the one-dimensional
PDF for the matter density parameter $\Omega_0$.}}
\end{figure}
\end{center}

\section{Conclusions}

In this work we have investigate the Rastall's proposal of modification of General
Relativity with respect to the problem of structure formation. This proposal is
equivalent to change the usual conservation law of General Relativity. In some
sense, it can be seen as a modification of the equation of state of a given fluid
from the dynamical point of view. In this sense, the Rastall's theory can be interesting
in order to obtain an accelerated expansion of the universe without introducing exotic
fluids. For example fluids with positive or null pressure may induce a dynamics typical
of fluids of negative pressure. This behaviour has already been remarked, for example,
in the Brans-Dicke theory \cite{rose}.
\par
Our results indicate that the Rastall's theory faces many problems at perturbative level.
Considering a homogenous and isotropic universe for example, the effective equation of
state of the background is entirely reproduced at perturbative level, leading to high instabilities when this effective equation of state implies negative pressure. One way
out is to consider a two-fluid model, one of the fluids obeying the usual conservation
law and the other one following the Rastall's prescription. The results point out to
a configuration which still reduces the Rastall's theory to General Relativity.
\par
When we keep the framework of two components which obeys different conservation laws,
but with a self-interacting scalar field playing the role of the ``Rastall's fluid'',
more interesting features appear. In particular, the Rastall's proposal can be seen as
a modification of the Klein-Gordon equation, similarly to what happens in the K-Essence
theories \cite{mukhanov}. Again the configuration corresponding to General Relativity
is favored, but other configurations with $\gamma \sim 0$ are also possible even if some statistical subtleties appear. All these considerations seem
to indicate to the specificity of General Relativity and rule out the Rastall's theory.
\par
 A possible way out to save the Rastall's proposal is to consider the modification of
the usual conservation law as a manifestation of quantum effects (like particle production)
in the spirit of reference \cite{ademir}. The particle production in an expanding universe
may lead to new terms in the usual conservation law. Such possibility remains to explore.
In any case, it seems clear that structure formation asks for a fluid we behaves in
the background and at perturbative level with zero effective pressure (otherwise no
mass agglomeration can effectively occur), and this poses a serious problem in
the original framework of the Rastall's theory. In some sense, this has already been
remarked in reference \cite{capone}, forcing the authors of that work to use just one
fluid to fit the supernova type Ia data. But, such procedure seems to be impossible concerning
the structure formation problem.
\par
Other observational methods, like CMB, may be used to constrain better the Rastall's theory. But, the
results obtained in the present work seem to us to be strong enough to point out the difficulties that the
Rastall's theory face when confronted with observational data.
\par
\noindent
{\bf Acknowledgements:} We thank Oliver Piattela, Winfried Zimdahl and Hermano Velten for their
remarks and criticisms on the text. We thank also
CNPq (Brazil) and FAPES (Brazil) for partial financial support.

\end{document}